%% file: kpimunu_paper.tex
\documentclass[]{elsart}

\usepackage{graphicx}% Include figure files
\usepackage{dcolumn}%  Align table columns on decimal point
\usepackage{amsmath}              %% New math environments, commands.
\usepackage{amssymb}              %% New math symbols
\usepackage[]{fontenc}
\usepackage{longtable}

\begin{document}
\begin{frontmatter}
\title{Analysis of the semileptonic decay 
$D^0 \rightarrow \overline{K}{}^0 \pi^-\mu^+\nu$}
\input{authors}

\date{\today}
%%%%%%%%%%%%%%%%%%%%%%%%%%%%%%%%%%%%%%%%%%%%%%%%%%%%%%%%%%
\newcommand{\rvval}{1.706}
\newcommand{\rvstat}{0.677}
\newcommand{\rvsyst}{0.342}

\newcommand{\rtwoval}{0.912}
\newcommand{\rtwostat}{0.370}
\newcommand{\rtwosyst}{0.104}

\newcommand{\aval}{0.347}
\newcommand{\astat}{0.222}
\newcommand{\asyst}{0.053}

\newcommand{\brval}{0.337}
\newcommand{\brstat}{0.034}
\newcommand{\brsyst}{0.013}
\newcommand{\rv}{$r_v$}
\newcommand{\rtwo}{$r_2$}
\newcommand{\rthree}{$r_3$}

%%%%%%%%%%%%%%%%%%%%%%%%%%%%%%%%%%%%%%%%%%%%%%%%%%%%%%%%%%

\input{abstract}
\end{frontmatter}

\newcommand{\kine}{$\cos\theta_V~{\rm vs.}~\cos\theta_\ell~{\rm vs.}~q^2$}
\input{paper_voting}

\section{Acknowledgements}
\input{acknowledgements}

\bibliographystyle{myapsrev}
\bibliography{main}

\input{bib.tex}
%\clearpage

\end{document}

%% file: authors.tex
The FOCUS Collaboration\footnote{See http://www-focus.fnal.gov/authors.html for 
additional author information.}
\author[ucd]{J.~M.~Link},
\author[ucd]{P.~M.~Yager},
\author[cbpf]{J.~C.~Anjos},
\author[cbpf]{I.~Bediaga},
\author[cbpf]{C.~G\"obel},
\author[cbpf]{A.~A.~Machado},
\author[cbpf]{J.~Magnin},
\author[cbpf]{A.~Massafferri},
\author[cbpf]{J.~M.~de~Miranda},
\author[cbpf]{I.~M.~Pepe},
% ...

\address[ucd]{University of California, Davis, CA 95616}
\address[cbpf]{Centro Brasileiro de Pesquisas F\'{\i}sicas, Rio de Janeiro, RJ, Brasil}
\address[cinv]{CINVESTAV, 07000 M\'exico City, DF, Mexico}
\nopagebreak
\address[cu]{University of Colorado, Boulder, CO 80309}
\nopagebreak
\address[fnal]{Fermi National Accelerator Laboratory, Batavia, IL 60510}
\address[fras]{Laboratori Nazionali di Frascati dell'INFN, Frascati, Italy I-00044}
\address[ugj]{University of Guanajuato, 37150 Leon, Guanajuato, Mexico}
\address[ui]{University of Illinois, Urbana-Champaign, IL 61801}
\address[iu]{Indiana University, Bloomington, IN 47405}
\address[korea]{Korea University, Seoul, Korea 136-701}
\address[kp]{Kyungpook National University, Taegu, Korea 702-701}
\address[milan]{INFN and University of Milano, Milano, Italy}
\address[nc]{University of North Carolina, Asheville, NC 28804}
\address[pavia]{Dipartimento di Fisica Nucleare e Teorica and INFN, Pavia, Italy}
\address[pr]{University of Puerto Rico, Mayaguez, PR 00681}
\address[sc]{University of South Carolina, Columbia, SC 29208}
\address[ut]{University of Tennessee, Knoxville, TN 37996}
\address[vu]{Vanderbilt University, Nashville, TN 37235}
\address[wisc]{University of Wisconsin, Madison, WI 53706}

%% file: abstract.tex
\begin{abstract}
{\normalsize Using data collected by the fixed target Fermilab experiment FOCUS, 
we present several first measurements for the semileptonic decay 
$D^0 \rightarrow \overline{K}\,\!^0\pi^-\mu^+\nu$. Using a model that includes
a $\overline{K}\,\!^0 \pi^-$ S-wave component, we measure the form factor ratios to
be 
$r_v=\rvval \pm \rvstat\pm \rvsyst$
and
$r_2=\rtwoval\pm \rtwostat\pm \rtwosyst$
and the S-wave  amplitude to be
$A=\aval \pm \astat\pm \asyst~{\rm GeV}^{-1}$.
Finally, we measure the vector semileptonic branching ratio
$\frac{\Gamma(D^0 \rightarrow  K^{*}(892)^{-}\mu^+\nu)}{\Gamma(D^0 \rightarrow \overline{K}\,\!^0\pi^-\pi^+)}=
\brval \pm \brstat\pm \brsyst$.
} 
{\normalsize \par}
\end{abstract}

%% file: paper_voting.tex
\section{Introduction}

Cabibbo allowed semileptonic decays have relatively large branching fractions 
and can be easily selected to achieve low levels of background
contamination. The experimental results can be directly compared to theory where 
decay rates are calculated from first principles and include QCD effects in
the form factors.
Form factors are predicted in several models 
(quark models~\cite{Scora:1995ty}, Lattice QCD~\cite{Bowler:1995bp}, and sum rules~\cite{Ball:1991bs}). 
Once the form factors are determined, the CKM matrix elements can be calculated. 
While there have been several measurements of the 
$D^+$~\cite{Link:2002wg,Adamovich:1998ia,Aitala:1997cm,Aitala:1998xu,Frabetti:1993jq,Kodama:1992tn}
form factors,  there are still no measurements of the $D^0$ form factors
for the vector semileptonic decays. 
We present the first 
measurement of the $D^0$ semileptonic form factor ratios for vector channels
and the branching ratio 
$\Gamma(D^0 \rightarrow  K^{*}(892)^{-}\mu^+\nu)/\Gamma(D^0 \rightarrow \overline{K}\,\!^0\pi^-\pi^+)$.\footnote{Charge 
conjugation is implied throughout this letter.} 
Furthermore, we present an investigation of the S-wave component of the $\overline{K}\,\!^0\pi^-$ system 
and the measurement of its amplitude.
This S-wave representation~\cite{Link:2002ev} was first used by FOCUS for the analysis of the decay 
$D^+ \rightarrow K^- \pi^+\mu^+\nu$.

\begin{figure}[!]
\begin{center}\includegraphics[ width=6.5cm, height=4.5cm ]{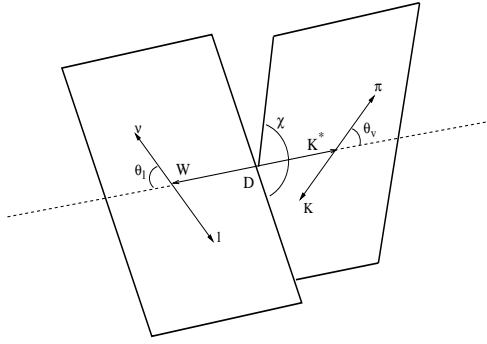}\end{center}
\caption{\label{fg:decayscheme} Schematic of the decay $D \rightarrow K^*\ell^+\nu$ for angular
variables definition.}
\end{figure}

The four--body decay amplitude can be parameterized by two masses and three angles. 
We use $M(\overline{K}\,\!^0\pi^-)$, $q^2=(P_\mu+P_\nu)^2$, and the three angles defined in
Fig.~\ref{fg:decayscheme}: $\cos\theta_V$ (the angle between the $\pi$ and the $D$ in the $K^*$ 
rest frame),  $\cos\theta_\ell$ (the angle between the $\nu$ and the $D$ in the $W$
rest frame), and
$\chi$ (the angle between the decay planes of the $K^*$ and the $W$).

With these definitions the decay amplitude is written 
as:\footnote{This model assumes that the $q^2$ dependence of the S-wave amplitude coupling to the
virtual $W^+$ is the same as the $H_0$ helicity amplitude describing the $K^*$ component. A study 
with as much as 100 times the  statistics of this analysis has been performed for the FOCUS 
analysis of the decay $D^+ \rightarrow K^- \pi^+ \mu^+ \nu$~\cite{Link:2002ev}. This study, 
where a significantly
different kinematic dependence for the S-wave has been used, has reported a change in the
form factors of less than 6\% of the statistical error.}
\begin{align}
\label{eq:ampli2}
&\frac{d^5\Gamma}{dm_{K\pi} dq^2 d\cos\theta_v d\cos\theta_\ell d\chi}\propto
 K(q^2-m^2_\ell)\\
&
\nonumber
\left \{ \begin{vmatrix}
(1+\cos\theta_\ell)\sin\theta_v e^{i\chi}B_{K^{*-}}H_+\\
-(1-\cos\theta_\ell)\sin\theta_ve^{-i\chi}B_{K^{*-}}H_-\\
-2\sin\theta_\ell (\cos\theta_vB_{K^{*-}}+Ae^{i\delta})H_0 
\end{vmatrix}^2 
+\frac{m^2_\ell}{q^2}\begin{vmatrix}
\sin\theta_\ell \sin\theta_v B_{K^{*-}}(e^{i\chi}H_+ +e^{-i\chi}H_-)\\
+2\cos\theta_\ell(\cos\theta_vB_{K^{*-}}+Ae^{i\delta})H_0\\
+2 (\cos\theta_vB_{K^{*-}}+Ae^{i\delta})H_t 
\end{vmatrix}^2 \right \}
\end{align}

where $K$ is the $K \pi$ system momentum in the $D$ rest frame and 
$B_{K^{*-}}$ and $Ae^{i\delta}$ are the Breit-Wigner and the S-wave components describing the 
spin one and spin zero states of $\overline{K}\,\!^0\pi^-$, respectively.
The four form factors ($A_{1,2,3}$ and $V$) that are necessary to describe a decay 
$D \rightarrow V \ell \nu$ (where $V$ stands for Vector), are included in the four helicity
amplitudes:
\begin{align}
%\nonumber
&H_\pm(q^2)=[(M_D+m_{K\pi})A_1(q^2)\mp\frac{2M_DK}{M_D+m_{K\pi}}V(q^2)]\\
%\nonumber
%
%&H_0(q^2)\!=\!\frac{1}{2m_{K\pi}\sqrt{q^2}}[(M^2_D\!-\!m^2_{K\pi}\!-\!q^2)(M_D\!+\!m_{K\pi})A_1(q^2)
%\!-\!\frac{4M_D^2K^2}{M_D+m_{K\pi}}A_2(q^2)]\hspace{-0.66cm}\\
%
\nonumber
&H_0(q^2)=\frac{1}{2m_{K\pi}\sqrt{q^2}}[(M^2_D - m^2_{K\pi} - q^2)(M_D + m_{K\pi})A_1(q^2)\\
 &\hspace{4cm}- \frac{4M_D^2K^2}{M_D+m_{K\pi}}A_2(q^2)]\\
\nonumber
&H_t(q^2)=\frac{M_DK}{m_{K\pi}\sqrt{q^2}}[(M_D+m_{K\pi})A_1(q^2)-
\frac{M^2_D-m^2_{K\pi}+q^2}{M_D+m_{K\pi}}A_2(q^2)\\
&\hspace{4cm}+\frac{2q^2}{M_D+m_{K\pi}}A_3(q^2)]
\end{align}

A pole mass form is assumed for the form factors:
\begin{align}
%\nonumber
A_i(q^2)=\frac{A_i(0)}{1-\frac{q^2}{M_A^2}} \hspace{2cm} V(q^2)=\frac{V(0)}{1-\frac{q^2}{M_V^2}}
\end{align}

\noindent
where $M_A$ and $M_V$ are the masses of the lowest $c \overline s$ states with the same quantum numbers as
the $W$, namely  $M_A=2.5$~GeV/$c^2$ and $M_V=2.1$~GeV/$c^2$ (which correspond to the masses of 
the $D_{s1}^{+}$ and $D_s^{*+}$, respectively). 
By including the parameter $A_1(0)$ in the constant that multiplies Eq.~\ref{eq:ampli2}, the decay amplitude
can be parameterized by the ratios of the form factors at $q^2=0$:

\begin{equation}
%\nonumber
r_v=\frac{V(0)}{A_1(0)},\hspace{1cm}r_2=\frac{A_2(0)}{A_1(0)},\hspace{1cm}r_3=\frac{A_3(0)}{A_1(0)}\\
\end{equation}

We measure $r_v$ and $r_2$. We have inadequate sensitivity to determine $r_3$ and we 
set its value to zero. From variations of this value we determine that the systematic uncertainty 
from setting $r_3=0$ is negligible.

\section{Event Reconstruction and Selection}

FOCUS is a photoproduction experiment which collected data during the
1996--1997 fixed-target run at Fermilab. 
The experiment, which is an upgrade of Fermilab experiment E687~\cite{Frabetti:1990au,Frabetti:1992bn}, 
is characterized by excellent vertex resolution and particle identification. 
For about 2/3 of the data taking the experimental target
was interleaved with a target silicon system~\cite{Link:2002zg}.
The track reconstruction
downstream of the target is performed by four stations of silicon microstrips (SSD) and five stations of
proportional wire chambers.
The momentum of charged tracks is measured by the deflection in two magnets of opposite polarity.
Charged particle identification is performed by three multi-cell threshold \v{C}erenkov
counters for electrons,
pions, kaons, and protons~\cite{Link:2001pg}. Combining the information on the track momentum and the
number of
photoelectrons produced in the cells inside the $\beta=1$ cone, a 
negative log-likelihood  variable ($W$)
for the hypothesis of the particle to be an electron, pion, kaon, or proton is determined.
Particle identification is performed by a comparison of the probabilities for the different hypotheses and
by requiring the hypothesis for the candidate particle to be higher than for the other hypotheses.
Muons are identified by the hits left in tracking systems after
penetrating approximately 21 interaction lengths of shielding
material \cite{Link:2002ev}.

In reconstructing $D^0 \rightarrow \overline{K}\,\!^0\pi^-\mu^+\nu$, we select combinations of
two charged tracks of opposite sign where one is identified as a pion and the other as a muon. 
For pion identification we require the pion hypothesis not to be disfavored by more than six units of
log-likelihood compared to the hypothesis with highest confidence level 
(min(${\it W})-{\it W}_\pi>-6$), and to be favored by one unit 
of log-likelihood over the kaon hypothesis (to reduce
the contamination from $D^0 \rightarrow K^-\mu^+\nu$). 
For  muon identification we require the track to have been reconstructed in the muon system 
(with at most one plane missed) with a confidence level greater than 1\%. 
In order to reject background from the decays $\pi^+ / K^+ \rightarrow \mu^+ \nu$,
we require  the muon trajectory to be consistent
through the two analysis magnets with a confidence level greater than 1\%. 
Each track must have momentum greater than 10~GeV/$c$. 

The two tracks are used to  form the $D^0$ decay vertex, which is required to have C.L.$>5$\%, where C.L. is
the confidence level.
To reduce contamination from higher multiplicity decays, we require the probability for any
other track reconstructed in the SSD system to come from the decay vertex to be lower than 0.1\%. This
requirement does not apply to the tracks used for the primary vertex reconstruction. To minimize background
from hadronic re-interactions in the target, the decay vertex must lie at least
one sigma outside of the target. 
The $\overline{K}\,\!^0$ is reconstructed as a $K_S^0$ from the decay 
$K_S^0 \rightarrow \pi^-\pi^+$~\cite{Link:2001dj}. 
The invariant mass is required to be within three sigma of
the nominal $K_S^0$ mass. 
If the pions are reconstructed using information from the silicon
vertex detectors, the reconstructed $K^0_S$ direction is used
in the reconstruction of the $D^0$ vertex.
In order to enhance the probability that our $K_S^0\pi^-$ combination
comes from a $K^{*}(892)^{-}$, we require the reconstructed $K_S^0\pi^-$ mass to
be within one $\Gamma$ of the nominal $K^{*}(892)^{-}$ mass.
The $K^{*}(892)^{-}$ natural width $\Gamma$ (50~MeV/$c^2$) is much larger than the 
experimental resolution on the reconstructed $K_S^0\pi^-$ mass (5~MeV/$c^2$).
The invariant
mass $M(K_S^0\pi^-\mu^+)$ is required to be lower than 1.8~GeV/$c^2$. 
This cut significantly reduces combinatoric background since $M(K_S^0\pi^-\mu^+)$ is
kinematically limited to be below the nominal $D^0$ mass and rejects
$D^0 \rightarrow K_S^0\pi^-\pi^+$ decays when one of the pions is misidentified as a muon.

We use the SSD tracks which have not been used in the $D^0$ decay reconstruction to form 
primary vertex candidates.
Each candidate is formed by starting with two tracks that make a vertex with C.L.$>1$\%
and adding other tracks so
long as the C.L. remains greater than 1\%. When a vertex is formed the remaining tracks are used to form a second
candidate in the same way and so on for the other candidates. We select the candidate with the highest
multiplicity,  and arbitrate ties by
keeping the one with higher significance of separation from
the secondary vertex. The significance of separation, which is given by the ratio of the distance between
the  two vertices divided by its error ($\ell/\sigma_\ell$), is required to satisfy $\ell/\sigma_\ell>5$.
We ``tag'' the $D^0$ by requiring that it comes from the decay $D^{*+} \rightarrow D^0 \pi^+_s$. 
The kinematics of this decay result in the pion having low momentum and being called a ``slow'' pion
($\pi_s$).
The pion must be one of the tracks used in the primary vertex reconstruction. It must have 
min($W)-W_\pi>-6$ and $p>2$~GeV/$c$.

\section{Fitting for the Form Factor Ratios and $\overline{K}\,\!^0\pi^-$
S-wave Amplitude}

In order to determine the form factor ratios $r_v$ and $r_2$ we use a combined fit of the mass difference
$\Delta M=M(D^*)-M(D)$ and the 
three  dimensional  distribution  \kine.
For the $\Delta M$ component we use 60 bins  
in the region 0.14--0.20~GeV/$c^2$. For the \kine~distribution we
select events with $ \Delta M < 0.15$~GeV/$c^2$, and divide
the  phase space into four equally spaced bins for each of the two angular 
variables and two equally spaced bins for  $q^2$. The  $\Delta M$ distribution, 
where signal and background events have a
very different shape, is used to evaluate the background level.
The binning choice for \kine~
gives information on the angular
distributions of the $W$ and the $\overline{K}\,\!^0 \pi^-$ decays for two regions of $q^2$. At low $q^2$ the
angular dependence is more dramatic, while a more isotropic behavior is expected for high $q^2$ values,
where the helicity amplitudes contribute with similar strength.

Two methods are used to find the momentum of the missing neutrino.
To compute the $q^2$ we use a ``$D^*$ cone'' algorithm. By imposing energy and momentum conservation
in the $K^{*}\mu$ rest frame and by constraining the $D$ and the $D^*$ to their nominal masses, the
magnitude of $p(D^0$) (which in this frame is equal to $p(\nu$)) is determined, 
but the direction lies on a cone. 
The direction is chosen by selecting the solution that gives the 
best $\chi^2$ when compared to the $D^0$ direction as given by the line connecting the two vertices.
To compute the mass difference,  
we determine the neutrino momentum using the ``neutrino closure'' algorithm.  
This method is based on energy and momentum conservation for the decay
$D^0 \rightarrow \overline{K}\,\!^0 \pi^- \mu^+ \nu$ and uses the nominal  mass of the $D^0$ meson. 
The algorithm allows us to determine the neutrino momentum
up to a two fold ambiguity, 
which is resolved by choosing the solution with lowest $\Delta M$. 
Monte Carlo studies show that this choice is most often the correct solution.

We use a binned maximum  likelihood fitting technique with:
\begin{equation}
\mathcal {L}=\prod_{ijk}\frac{n_{ijk}^{s_{ijk}}e^{-n_{ijk}}}{s_{ijk}!} \times \prod_{m}\frac{N_{m}^{S_{m}}e^{-N_{m}}}{S_{m}!}
\end{equation}
where $s_{ijk}$ ($n_{ijk}$) is the number of  observed (expected) events
in the ${ijk}^{th}$ bin of the three dimensional
distribution and $S_{m}$ ($N_{m}$) is the number of  observed (expected) events
in the $\Delta M$ distribution. The number of expected events is given by signal and background contributions.
Non-charm backgrounds are essentially removed by the $\ell / \sigma_\ell$ requirement, by discarding
events where the reconstructed decay vertex of the $D^0$ lies within one standard deviation from
the target, and by the muon requirement.
Contamination from charm decays is accounted for by using a Monte Carlo 
(that will be called MC$_\mathrm{BKG}$) which simulates all known charm decays other than our signal mode. The shapes
for both distributions are taken from the distributions of the reconstructed events in  
MC$_\mathrm{BKG}$, and their amplitudes are  free to float. The  background levels 
in the two distributions are tied by imposing
that the yield of the MC$_\mathrm{BKG}$ in the \kine~distribution is
equal to the area of the background shape in the $\Delta M$ distribution for 
$\Delta M< 0.15$~GeV/$c^2$. This corresponds to the selection cut imposed on the events in the 
\kine~distribution.
For the $\Delta M$ distribution,  the signal shape is taken from Monte Carlo generated 
$D^0 \rightarrow \overline{K}\,\!^0\pi^-\mu^+\nu$ events.
For \kine~ the signal contribution to $n_{ijk}$ 
is computed as the number of events
generated in the bin $ijk$ corrected by the efficiency for that bin. 
We calculate the generated number of event 
in  bin ${ijk}$ as a function of the fit parameters $r_v$ and $r_2$ using a Monte Carlo event
weighting procedure based on Ref.~\cite{Schmidt:1992si}.
For each Monte Carlo event generated in the bin ${ijk}$, we fill that bin with a weight given
by the ratio of the decay amplitude in Eq.~\ref{eq:ampli2} evaluated for the fit parameters
$r_v$ and $r_2$ over the decay amplitude evaluated for the input Monte Carlo 
values.\footnote{The FOCUS
Monte Carlo simulation uses the $D^+ \rightarrow K^-\pi^+\mu^+\nu$ form factor ratios and the 
S-wave parameters measured 
in~\cite{Link:2002wg}:
$r_v=1.504\pm 0.057\pm 0.039$,
$r_2=0.875\pm 0.049\pm 0.064$,
$A=0.330\pm 0.022\pm 0.015$, and $\delta=0.68\pm0.07\pm0.05$.} The signal yields in the 
\kine~and in the $\Delta M$ distributions are constrained in the
same way as explained for the background events.
\begin{figure}[!]
\begin{center}\includegraphics[width=14cm]{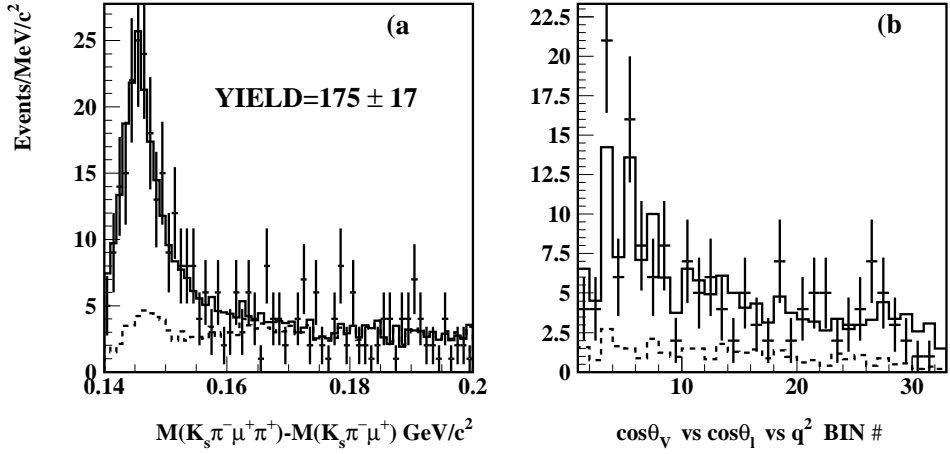}\end{center}
\caption{\label{Fg:fits} a) $\Delta M$ fit, b) \kine~fit. Points
with error bars are data, histogram is the fit, dashed line is the background component.}
\end{figure}

The combined fit is shown in Fig.~\ref{Fg:fits}. The number of signal events is 
$175\pm17$. 
The $\chi^2$ per degree of freedom in
Fig.~\ref{Fg:fits}b is 32/27 which 
corresponds to a confidence level of 22\%. We measure the form   factor ratios to be:
\begin{align}
 r_v &=  1.706 \pm 0.677 \\
 r_2 &=  0.912  \pm  0.370 
\end{align}

where the errors are statistical.

The fit for the amplitude of the S-wave is performed with the same technique as the fit for 
the form factor ratios. We fix the form factor ratios to the values found above and, 
based on isospin symmetry, we fix the phase to  0.68,
the value measured for the $D^+ \rightarrow K^-\pi^+\mu^+\nu$ decay.
As described in section~\ref{sec:system},  the possible bias due to this assumption
is included in the systematic uncertainty evaluation. We find that $A$ does not depend strongly on the 
phase. We measure:
\begin{align}
&A =  0.347 \pm  0.222~{\rm GeV}^{-1}
\end{align}
where the error is statistical.

\section{The Branching Ratio $\Gamma(D^0 \rightarrow K^{*}(892)^{-}\mu^+\nu)/\Gamma(D^0 \rightarrow
\overline{K}\,\!^0 \pi^-\pi^+)$}
The branching ratio is measured by dividing the efficiency corrected yields of the two modes.
The normalization mode $D^0 \rightarrow \overline{K}\,\!^0 \pi^-\pi^+$ is reconstructed following the same
procedure and applying the same requirements as for the 
$D^0 \rightarrow \overline{K}\,\!^0 \pi^-\mu^+\nu$ mode (when
possible), in order to minimize bias due to possible inaccuracies in the Monte Carlo evaluation of the 
efficiency for the
$K_S^0$, which is reconstructed in a very different way from ordinary tracks. 
$D^{*+}$ tag, vertex reconstruction, $K_S^0$ reconstruction, and particle identification 
(except  muon identification) 
are the same as for the semileptonic mode. The $\pi$ with the opposite charge of the $\pi_s$  must pass identical requirements
as the $\pi^-$ in the semileptonic mode. 
The trajectory of the $\pi$ with the same charge as the $\pi_s$ must be
consistent through the two analysis magnets, as we require for the $\mu^+$.
In addition it is required to have 
min$(W)-W_\pi>-6$. For the hadronic mode we do not require the event to be in the
mass window around the $K^{*}(892)^{-}$ nominal mass. The invariant mass $M(K_S^0\pi^-\pi^+)$ must lie within 
24~MeV/$c^2$ of the fit $D^0$ mass,
both for data and Monte Carlo. This window corresponds to a two sigma cut.
The $\Delta M$ distribution is fit to
 two Gaussian distributions for the signal (in order to account for different resolutions)
and the following threshold function for the background:
\begin{equation}
{\rm BKG}(\Delta M)=a~(\Delta M-m_\pi)^{1/2}+b~(\Delta M-m_\pi)^{3/2}+c~(\Delta M-m_\pi)^{5/2}
\label{eq:BKG}
\end{equation}

\begin{figure}[!]
\begin{center}\includegraphics[width=14cm]{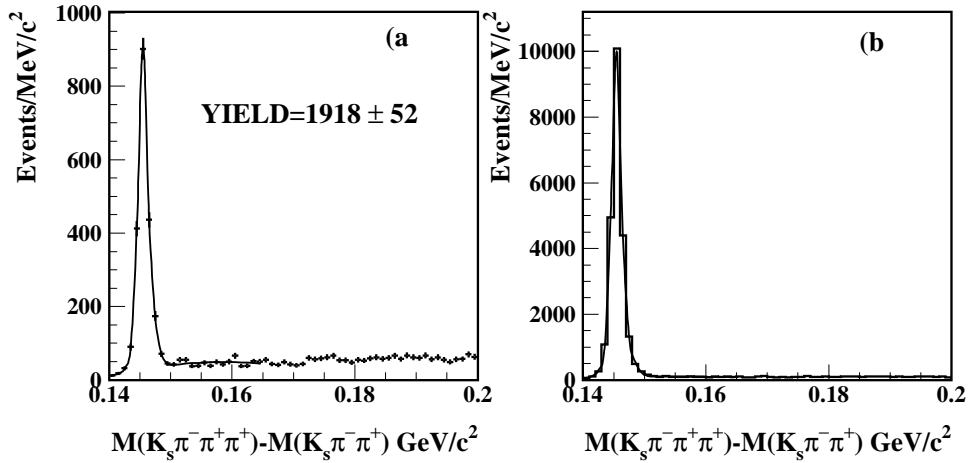}\end{center}
\caption{\label{Fg:fit_kpp} $\Delta M$ fit for data (a) and  Monte Carlo (b).
The signal is fit to two Gaussian distributions, the
background is fit to the threshold function in Eq.~\ref{eq:BKG}.}
\end{figure}

The fit for  data and Monte Carlo are shown in Fig.~\ref{Fg:fit_kpp}.
The yield from the fit to the data is $1918\pm52$ events. 
 
The efficiency corrected yield of $D^0 \rightarrow K^{*}(892)^{-}\mu^+\nu$ is determined by
correcting the efficiency corrected yield of $D^0 \rightarrow \overline{K}\,\!^0 \pi^- \mu^+\nu$ 
for the   amount of $K^{*-}$ in the $\overline{K}\,\!^0 \pi^-$ system. 
Since the  Monte Carlo simulation uses the form factor ratios and S-wave parameters measured in the
much higher statistics FOCUS analysis of $D^+\rightarrow K^-\pi^+\mu^-\nu$, 
which  are in  excellent agreement with that measured in the present analysis, 
we estimate that the correction factors for the number
of reconstructed events in data and Monte Carlo are the same, and therefore cancel out.
The number of generated events in Monte Carlo (which is used to calculate the efficiency) must be corrected
by the relative branching ratio 
$\Gamma(D^0 \rightarrow K^{*}(892)^{-} \mu^+\nu)/\Gamma(D^0 \rightarrow \overline{K}\,\!^0 \pi^-\mu^+\nu)$ 
used for the Monte Carlo simulation.
This can be evaluated by integrating over phase space the decay amplitude 
when the exclusive $K^{*}$ mode is generated, and dividing by the integral over phase space of 
the decay amplitude for the inclusive mode. We calculate that in our Monte Carlo simulation the
branching ratio is 0.95. 
We measure the branching ratio to be:
\begin{equation}
\frac{\Gamma(D^0 \rightarrow K^{*}(892)^{-}\mu^+\nu)}{\Gamma(D^0 \rightarrow
\overline{K}\,\!^0\pi^-\pi^+)}=0.337\pm0.034
\end{equation}
where the error is statistical.

\section{Systematic Uncertainty Evaluation}\label{sec:system}
We carefully considered and evaluated many possible
sources of systematic uncertainty in our results.
Systematic bias
can be generated by a poor Monte Carlo simulation of the detector performance, resulting in erroneous
estimation of the efficiency. Also the particular choice for the fitting technique and parameters
may bias the measurement. 

The accuracy and correct estimation of the errors reported by the
fitting method is evaluated by repeating the measurement on a thousand samples obtained from
fluctuating the bin entries in the data histogram. From the Gaussian distribution of the returned
values for each measured quantity (form factor ratios, S-wave amplitude, signal and
background
yields), we conclude that the fit method is not affected by systematic bias and returns correct
values for the errors. 

The Monte Carlo evaluation of the efficiency is investigated by repeating the measurements for
different variations of the selection cuts. As expected, when the efficiency is correctly
estimated (for our level of accuracy), the results are always stable within errors. 
We evaluate a possible bias due to the Monte Carlo simulation with the ``split sample''
technique, derived from the S--factor method used by the Particle Data Group~\cite{Eidelman:2004wy}. 
The data is split into statistically independent samples; for example, if the momentum
simulation is being investigated, the data is split into distinct momentum regions. 
The measurement is performed on each sample for the observable $x$ (e.g. $r_v$) and a $\chi^2$ for the 
hypothesis that the independent measurements are consistent is calculated. 
A poor consistency might result from a badly estimated efficiency with respect to the momentum. We
define poor consistency to be the case where $\chi^2>1$. In this case, the errors on the different
measurements are scaled in order to return $\chi^2=1$, and we calculate a systematic uncertainty for
the $x$ measurement by subtracting in quadrature the statistical error from the scaled error on
the weighted average of the independent measurements. 
Additional details are given in Ref.~\cite{Link:2002hi}.

The bias from fitting choices is evaluated as the variance of measurements obtained by
varying such choices. We vary the bin size both for the $\Delta M$ and the 
\kine~distributions. For $\Delta M$, we also vary the
fitting range. The $r_v$ and $r_2$ parameters are also evaluated
setting the S-wave parameters to zero. The S-wave amplitude is evaluated for two additional values
of the phase (at plus and minus one sigma from the reference value).
For the $r_v$, $r_2$, and $A$  fits, we include a
variation  on the fitting technique. This second fitting technique  
accounts for the efficiency in a different way.
The efficiency is
taken into account by using the weighting method on the reconstructed Monte Carlo events, instead
of the generated events. For each event that passes all the selection cuts, the bin in which the
event was generated in is filled with the weight described in Section 3.

For the branching ratio measurement, we investigate  the bias due to Monte Carlo input
parameters by varying the form factor ratios and the S-wave values, and by varying
 the resonant structure of $\overline{K}\,\!^0 \pi^-\pi^+$. 
 Also, a less refined simulation of the hadronic trigger is investigated.
The systematic bias from the model used in the Monte Carlo is evaluated as the 
variance of the three
measurements found with these variations and the standard result.

The total systematic uncertainty is given by the sum in quadrature of the uncertainties from
the independent sources.
Table~\ref{system} summarizes the results of the systematic uncertainty 
evaluation for all of
the measurements. Including the systematic uncertainty we measure:
\begin{align}
&r_v=1.706\pm0.677~({\rm stat})\pm 0.342~({\rm sys}) \\
&r_2=0.912\pm0.370~({\rm stat})\pm 0.104~({\rm sys})\\
&A=0.347 \pm0.222~({\rm stat})\pm 0.053~({\rm sys})~{\rm GeV}^{-1}\\
&\frac{\Gamma(D^0 \rightarrow  K^{*}(892)^{-}\mu^+\nu)}{\Gamma(D^0 \rightarrow \overline{K}\,\!^0\pi^-\pi^+)}=
0.337 \pm 0.034~({\rm stat})\pm 0.013~({\rm sys})
\end{align}

\begin{table}[htb!]
\begin{center}

\caption{The systematic uncertainties from the Monte Carlo efficiency and acceptance evaluation, the
fitting condition, and total for $r_v$, $r_2$, $A$, and the branching ratio are shown. For the branching
ratio,
the systematic from the input parameters and trigger simulation in the Monte Carlo is also
evaluated.}
\vspace{0.5cm}
\label{system}

Systematic Error 

\begin{tabular}{|c | c | c | c| c | }
\hline
\hline
  Source            &$ \sigma(r_v$) & $\sigma(r_2$)& $\sigma(A)({\rm GeV}^{-1}$) & $\sigma({\rm BR}$)  \\ 
\hline
MC Simulation       &0.0	    &0.0         &0.0    	 &0.0	            \\
\hline
Fit                 &0.342	    &0.104 	   &0.053 	 &0.002       	    \\
\hline
Model               &--	            &-- 	   &-- 	         &0.013        	    \\
\hline
Total               &0.342  	    &0.104	   &0.053 	 &0.013 	    \\
\hline
\hline
\end{tabular}
\end{center}
\end{table}

\section{Conclusions}

We have presented an analysis of the semileptonic decay 
$D^0 \rightarrow \overline{K}\,\!^0\pi^-\mu^+\nu$ using FOCUS data. 
Using a model which includes a 
$\overline{K}\,\!^0 \pi^-$ S-wave component that interferes with the dominant
$K^{*}(892)^{-}$ state, we have  
measured for the first time the $D^0$ form factor ratios for vector channels
and the S-wave amplitude.
We also report
the first measurement of the branching ratio 
$\Gamma(D^0 \!\rightarrow \! K^{*}(892)^{-}\mu^+\nu)/
\Gamma(D^0 \!\rightarrow \!\overline{K}\,\!^0\pi^-\pi^+)$.

\vspace{0.5cm}
\begin{table}[htb!]
\begin{center}
\caption{The measurement of $r_v$, $r_2$, and $A$  presented in this letter are compared to the FOCUS
results for the decay $D^+\!\rightarrow \!K^-\pi^+\mu^+\nu$. We fix the S-wave phase to 0.68, 
the value measured for the $D^+$.}
\vspace{0.5cm}
\label{Tb:results}

\begin{tabular}{|c | c| c|}
\hline 
                     & $D^0\!\rightarrow \! \overline{K}\,\!^0 \pi^-\mu^+\nu$  &$D^+\!\rightarrow \!K^-\pi^+\mu^+\nu$\\
\hline
  $r_v $             & $1.706\pm0.677\pm 0.342$   &$1.504\pm 0.057\pm 0.039$ \\ 	     
  $r_2 $             & $0.912\pm0.370\pm 0.104$   &$0.875\pm 0.049\pm 0.064$ \\ 	     
  $A({\rm GeV}^{-1})$  & $0.347\pm0.222\pm 0.053$ &$0.330\pm 0.022\pm 0.015$\\
      
\hline
\hline
\end{tabular}
\end{center}
\end{table}

From isospin symmetry, the expected values of $r_v$, $r_2$, and $A$  can be directly
compared to the results of the FOCUS measurements for the decay 
$D^+ \rightarrow K^- \pi^+\mu^+\nu$, which uses the same model as the analysis presented
in this letter. 
We find excellent agreement with the  values for the $D^+$, see Table~\ref{Tb:results}.
We calculate that in our model, where the $\overline{K}\,\!^0 \pi^-$ system is given by 
a scalar and a vector component, the scalar fraction  is 6\%. 

The branching ratio value can also be estimated from the $D^+$ analysis
using isospin symmetry:
\begin{align}
%\nonumber
\frac{\Gamma (D^{0}\!\rightarrow \!K^{*-} \mu^+ \nu)}{\Gamma (D^{0}\!\rightarrow\! \overline{K}\,\!^0 \pi^- \pi^+)}&\!=\!
\frac{\tau (D^{0})}{\tau (D^{+})}\!\times \!\frac{\Gamma (D^{+}\!\rightarrow K^{*0}\!\mu^+ \nu)}
{\Gamma (D^{+}\!\rightarrow\! K^- \pi^+ \pi^+)}\!\times\!
\frac{ {\mathcal B} (D^{+} \!\rightarrow \!K^- \pi^+ \pi^+)}{{\mathcal B}(D^{0}\! \rightarrow \!\overline{K}\,\!^0 \pi^- \pi^+)}
\label{isospinestimate}
%\\
%\nonumber
%\\
%&=0.366\pm 0.034
\end{align}

\begin{figure}[!]
\begin{center}\includegraphics[ width=7cm, height=6cm ]{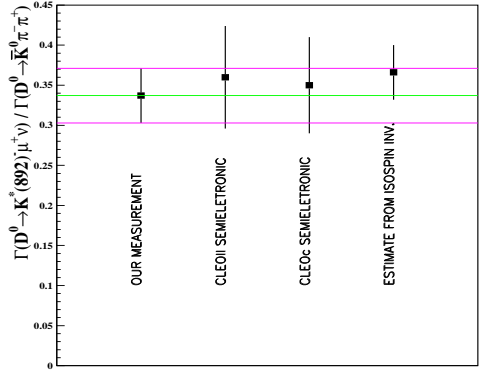}\end{center}
\caption{\label{Fg:comparebr} 
The $\Gamma(D^0 \rightarrow K^{*}(892)^{-}\mu^+\nu)/\Gamma(D^0 \rightarrow \overline{K}\,\!^0\pi^-\pi^+)$ 
FOCUS measurement is
compared to the CLEO-II measurement of the semielectronic mode
$\Gamma(D^0 \rightarrow K^{*}(892)^{-}e^+\nu)/\Gamma(D^0 \rightarrow \overline{K}\,\!^0\pi^-\pi^+$), 
with the CLEO-c preliminary measurement of $\mathcal {B}(D^0 \rightarrow K^{*}(892)^{-} e^+ \nu_e)$ divided by the Particle Data
Group average
for $\mathcal {B}(D^0 \rightarrow \overline{K}\,\!^0 \pi^-\pi^+)$, and to an estimate from isospin symmetry.
The semielectronic results are corrected to account for the smaller electron mass when compared to the
muon and they do not include the S-wave component.}
\end{figure}

Since the decay dynamics  do not depend on the lepton species, we compare the branching ratio result
to measurements that use the semielectronic channel.
Differences in the decay rate
are only due to the larger mass of the muon as compared to the electron. In the semimuonic mode the phase 
space is reduced and there is a more significant contribution from the $m^2$ term
of the decay amplitude (see Eq.~\ref{eq:ampli2}). According to the PDG, the electron values should be
corrected by a factor of 0.952 to compare to the muon results.
We apply this correction and compare our results to the CLEO-II measurement of 
$\Gamma (D^0 \rightarrow K^{*-} e^+ \nu)/\Gamma (D^{0} \rightarrow \overline{K}\,\!^0 \pi^-
\pi^+)$~\cite{Bean:1993zv}.
We also compare our results  to the recent preliminary result from CLEO-c of the absolute branching fraction 
$\mathcal {B}(D^0 \rightarrow K^{*-} e^+ \nu)$
(presented in conference proceedings~\cite{Gao:2004bw}) divided by the PDG average of
$\mathcal {B}(D^0 \rightarrow \overline{K}\,\!^0 \pi^-\pi^+)$. 
The comparison of our branching ratio measurement with the semielectronic results and with the
calculation in Eq.~\ref{isospinestimate} is shown in Fig.~\ref{Fg:comparebr}. Only the calculation
from isospin symmetry includes the effects of the S-wave component.
The three estimates come from different measurements, and are in excellent agreement with each other
and with our measurement. 

%% file: acknowledgements.tex
We wish to acknowledge the assistance of the staffs of Fermi National
Accelerator Laboratory, the INFN of Italy, and the physics departments of the
collaborating institutions. This research was supported in part by the U.~S.
National Science Foundation, the U.~S. Department of Energy, the Italian
Istituto Nazionale di Fisica Nucleare and Ministero dell'Universit\`a e della
Ricerca Scientifica e Tecnologica, the Brazilian Conselho Nacional de
Desenvolvimento Cient\'{\i}fico e Tecnol\'ogico, CONACyT-M\'exico, the Korean
Ministry of Education, and the Korean Science and Engineering Foundation.